\documentclass{svproc}
\usepackage{url}

\usepackage{amsmath, amssymb, mathrsfs, color, braket, mathtools, cite, float}
\usepackage[colorlinks=true]{hyperref}

\newcommand{\er}{\mathrm{err}}
\newcommand{\coh}{\mathrm{coh}}

\newcommand{\seq}{\mathrm{seq}}
\newcommand{\ind}{\mathrm{ind}}

\DeclarePairedDelimiter\ceil{\lceil}{\rceil}

\usepackage{cite}

\newcommand{\R}{\mathbb{R}}
\newcommand{\C}{\mathbb{C}}

\newcommand{\grp}[1]{\mathsf{#1}}
\newcommand{\spc}[1]{\mathcal{#1}}

\newcommand{\CPTP}{\mathsf{CPTP}}

\def\>{\rangle}
\def\<{\langle}

\newcommand{\map}[1]{\mathcal{#1}}
\newcommand{\tr}{\operatorname{tr}}

\newcommand{\St}{{\mathsf{St}}}

\newtheorem{prop}{Proposition}

\newtheorem{defi}{Definition}

\begin{document}
\mainmatter
\title{Fast tests for probing the causal structure of quantum processes}
\titlerunning{Fast tests of causal structure}  
\author{Giulio Chiribella\inst{1,2,3,4}  \and Swati\inst{1,5}}
\authorrunning{Chiribella and Swati} 
\tocauthor{Giulio Chiribella and Swati}
\institute{QICI Quantum Information and Compuation Initiative, Department of Computer Science, The University of Hong Kong, Pokfulam Road, Hong Kong\\
\email{giulio@cs.hku.hk}\\
\and
Department of Computer Science, University of Oxford, Oxford, OX1 3QD, United Kingdom\\
\and
Perimeter Institute for Theoretical Physics, Waterloo, Ontario N2L 2Y5, Canada\\
\and
HKU Shenzhen Institute of Research and Innovation, Kejizhong 2nd Road, Shenzhen, China
\and
Institute for Quantum Science and Engineering,  Department of Physics, Southern University of Science and Technology, Shenzhen, China\\
}

\maketitle              

\begin{abstract}
The identification of causal relations is a cornerstone of the scientific method.  Traditional approaches to this task  are based on classical statistics. However, such classical approaches do not apply in the quantum domain, where a broader spectrum of causal relations becomes accessible.  New approaches to quantum causal inference have been developed in recent years, and promising new features have been discovered. In this paper, we review and partly expand the framework and  results of Ref. \cite{Giulio2019}, which demonstrated  quantum speedups in the identification of various types of causal relations induced by  reversible processes. 

\keywords{causal inference, quantum estimation theory, indefinite causal order}
\end{abstract}
\section{Introduction}

Identifying  cause-effect relations  is a fundamental problem in science and engineering  \cite{spirtes2000causation, pearl2009, pearl2014}. In its simplest form, the problem can be described as follows:  An experimenter has  collected some amount of  raw data  about the values of a set of variables, and wants  to determine whether a certain variable $X$ influences another variable $Y$ in the set. Since the data are typically subject to noise and fluctuations, the problem of identifying causal  relations is ultimately a statistical inference problem. 

Traditionally,  causal inference  methods have  been designed for classical variables. Recent advances in quantum science and engineering motivate an extension of these methods to the quantum domain. Experimental techniques  can now address individual quantum systems, initialise them in a given quantum state, and subject them to a variety of  quantum measurements. In this context, the presence of causal dependencies between two  quantum systems acquire a concrete practical relevance. For example, if the interaction of two  quantum spins induces a causal dependence between them, then the state of one spin can be  controlled  by operating on the other spin, to some degree that is determined by the strength of the causal relation.

For quantum  systems, classical methods of causal inference turn out be inadequate. The reason is that such methods assume that randomness can always be reduced to ignorance about the initial conditions of some additional, latent variable. This assumption is at odds with the violation of Bell inequalities, which rules out local realistic models for quantum correlations.   For this reason, classical causal models cannot be applied to the Bell scenario \cite{wood2015lesson}.  A similar conclusion also holds for more general scenarios, including more than two quantum systems, and/or timelike separations \cite{chaves2018quantum, van2018quantum}. 

The breakdown of classical causal inference methods calls for a  new, genuinely quantum formulation of the notions of cause and effect.  Several frameworks have been proposed to date, with different features and sometimes different purposes.   
The framework of {\em quantum combs}, introduced by Chiribella, D'Ariano, and Perinotti in a series of works \cite{Chiribella-circuit2008, Chiribella-memory2008, chiribella-dariano-2009-pra }, describes networks of quantum processes connected with one another according to a given causal structure. In this framework, one can express the fact that a given process induces a causal dependence between a quantum system and another. More broadly, this notion can also be generalised to probabilistic theories beyond quantum mechanics \cite{Chiribella-probabilistic2010}, and to an even broader class of theories described by symmetric monoidal categories  \cite{coecke2012picturing, Coecke-terminality2014}.  
Frameworks for describing causal relations in quantum theory and beyond have been developed in \cite{henson2014theory, pienaar2015graph, costa2016quantum}.  
More recently, a quantum framework for describing causal relations between quantum systems has been developed by Allen {\em et al} \cite{allen2017quantum}. This framework, known as {\em quantum causal models}, can be regarded as an enrichment of the framework of quantum combs, with new conditions that allow one to express the {\em conditional  independences} among quantum variables.

Given a framework for describing causal relationships among quantum variables, one can develop strategies for identifying such relations. Two interesting examples  were presented in Refs \cite{fitzsimons2015quantum,ried2015}. In these works, the authors considered the problem of determining whether two quantum systems, say two photons, are in a spacelike or timelike configuration. Equivalently, this amounts to determining whether the states of the two systems are marginals of a bipartite quantum state (spacelike configuration, corresponding to a past common cause in the joint preparation of the two systems), or whether the state of one system is obtained from the state of the other by a quantum process (timelike configuration, corresponding to a cause-effect relation directed from one system to the other).  Remarkably, the authors found out that certain quantum correlations can distinguish between these two situations, thus defying the classical motto  ``correlation does not imply causation".  

 The fact that quantum correlations can be witnesses of causal relationships suggests that   quantum measurements could offer more powerful ways to identify causal relations compared to their classical counterpart.  In a recent work \cite{Giulio2019}, Chiribella and  Ebler showed that quantum features like entanglement and superposition can lead to speedups in the identification of various types of causal relations.  In particular, they addressed the following question: given a variable, which variable out of a list of candidate variables, is the effect of it? For simplicity, they focussed on the basic scenario where the cause-effect relation is induced by a reversible process. 
  Classically, it turns out that the  minimum probability of error  is  $p_{\rm err}^{\rm C} (N)  =  1/2d^{N-1}$, where $d$ is the dimension of the quantum systems associated to the given variables, and  $N$ is the number of times the variables are probed.  This means that the  error probability decays exponentially as $p_{\rm err}^{\rm C} (N) \approx  2^{  -  R_{\rm C} N}$, with decay rate $R_{\rm C}    =    \log_2  d$.   In stark contrast, Chiribella and Ebler showed that the error probability for quantum strategies decays quadratically faster, with an exponential decay rate equal to $2 \log_2 d$.   Similar advantages arise in the task of identifying the presence of a causal link between two variables, and in the task of identifying the cause of a given effect.

This paper reviews the framework and the results of Ref. \cite{Giulio2019} in a way that is suitable for non-specialists. The paper is organised as follows: in Section \ref{sec:prelims}, we  provide preliminary notions that will be used later in the paper. In Section \ref{sec:opt-class}, we review the problem of identifying of the effect of a given variable, and we provide the minimum  error probability and its decay rate for classical strategies.  Then, we provide quantum strategies that  achieve a  speedup over their classical counterpart (Section \ref{sec:opt-quant}). In Section \ref{sec:apps}, we generalize the above results for the case of multiple hypotheses and provide some applications. Finally, we conclude in Section \ref{sec:conc} by discussing    directions of future research.

\section{Preliminaries}
\label{sec:prelims}
In this section  we review and expand the framework of Ref. \cite{Giulio2019},  providing  some additional definitions that help clarifying the different types of causal relations induced by quantum processes.
\subsection{Notation}

For a given Hilbert space  $\spc H$, we denote by $ B(\spc H)$ the space of bounded operators on $\spc H$, by  $T(\spc H)$ the space of trace-class operators, and by $\St(\spc H)   :  =\{  \rho  \in T(\spc H)~|~ \rho \ge  0  \, , \,   \tr[\rho] =1\}$ the convex set of density operators.    In the context of causal inference, quantum systems  are often  called {\em quantum variables}.  
 We will denote quantum systems by Roman letters, such as $A, B, \dots$, and the corresponding Hilbert spaces as $\spc H_A, \spc H_B, \dots$, respectively.  We will use the shorthand notation $\St (A) : =  \St (\spc H_A)$.

 Let $A$ and $B$ be two quantum systems, and  let   $\spc H_A$ and $\spc H_B $ be the corresponding Hilbert spaces. We will denote by $A\otimes B$ the composite system consisting of subsystems $A$ and $B$, corresponding to the tensor product Hilbert space $\spc H_{A\otimes B}  =  \spc H_A\otimes \spc H_B$.  
The partial trace over the Hilbert space  $\spc H_A$ will be denoted as $\tr_A$. 
   
 A quantum process with input $A$ and output  $B$ is described by a linear,  completely positive, trace-preserving map  $\map C:  T  (\spc H_A)  \to T (\spc H_B)$, mapping input states $\rho  \in  \St(A)$ into output states $\map C  (\rho)  \in  \St (B)$.   Linear, completely positive and trace-preserving maps will be sometimes abbreviated as CPTP maps.   The set of CPTP maps from $T(\spc H_A)$ to $T (\spc H_B)$ will be denoted by $\CPTP (A\to B)$.

 \subsection{Cause-effect relations induced by quantum processes}

  Let $A$ and $B$ be two quantum systems.  
    
  \begin{defi}\label{def:causeeffectprocess}

We say that a process $\map C \in \CPTP (A\to B)$  {\em induces  a cause-effect relation} from $A$ to $B$ if and only if $\map C$ is non-constant.  
 If   this is the case, then we say that $A$ is a cause for $B$, and that $B$ is an effect of $A$.  
\end{defi}
In other words, a process induces a causal relation from $A$ to $B$  if and only if changing the state of system $A$ can induce  changes  in the state of system $B$.  The ability to induce changes serves as a witness of the fact that $A$ is a cause for $B$.  


Definition (\ref{def:causeeffectprocess}) can be generalized  to processes involving multiple inputs and outputs.        
\begin{defi}\label{def:causeeffectprocess1}
We say that a bipartite process $\map D\in \CPTP (  (A\otimes A' )\to (B\otimes B'))$  {\em induces  a cause-effect relation} from $A$ to $B$ if and only if there exists at least one  state $\alpha  \in  \St (\spc H_{A'})$ such that  the {\em reduced process}   
\begin{align}
\map D_\alpha  \in   \CPTP (A\to B)\, ,  \qquad  \map \rho   \mapsto  \tr_{B'}   [\map D   (\rho \otimes \alpha) ]      
\end{align} is non-constant.
 If this is the case, then we say that $A$  is a {\em cause} for $B$, and that $B$ is an {\em effect} of $A$.  
 \end{defi}
Note that a process $\map C \in \CPTP (A\to B)$ may  not induce a cause-effect relation from $A$ to $B$, but may still be  the reduced process of some other process $\map D\in \CPTP (  (A\otimes A' )\to (B\otimes B'))$  that does induce a cause-effect  relation from $A$ to $B$.  This observation shows that the presence of a cause-effect  relation, as defined in Definitions  (\ref{def:causeeffectprocess})   and (\ref{def:causeeffectprocess1}), is a property of the process under consideration, and not just of the variables $A$ and $B$.   In other words, a cause-effect relation that is actually present may not be detected by inspecting the process from $A$ to $B$ alone.  

The notion of ``process inducing a cause-effect relation''  provided in Definitions  (\ref{def:causeeffectprocess})   and (\ref{def:causeeffectprocess1})  is rather weak, because it allows the  influence of the cause on the effect to  be arbitrarily small.    A stronger notion arises when the  cause-effect relation is  {\em faithful}, in the following sense:

\begin{defi}  We say that a process $\map C\in \CPTP(A\to B)$  {\em induces  a  faithful   cause-effect relation} from $A$ to $B$ if and only if   $\map C$ is correctable,   meaning that there exists another process $\map R\in \CPTP(B\to A) $ such that 
\begin{align}  \map R \circ \map C  =  \map I_A \,,
\end{align} where $\map I_A \in  \CPTP  (A\to A)$ is the identity process.
When this is the case, we say that $B$ is a {\em causal intermediary} of $A$.  
\end{defi}  
Intuitively, variable $B$ is a causal intermediary for variable $A$ if  all the possible causal influences of $A$ can be reconstructed from  $B$. In other words,  every process from $A$ to  a third  variable $B'$  must factorise into the  process from $A$ to $B$, followed by  some process from $B$ to $B'$.  
This intuition is formalised by the following proposition:  
\begin{prop}
The process  $\map C\in \CPTP (A\to B)$  induces a faithful cause-effect relation from $A$ to $B$ if and only if, for every quantum system $B'$,  and for every process $\map E  \in  \CPTP (A\to B')$ there exists a process $\map D  \in  \CPTP (B\to B') $ such that 
\begin{align}
\map E   =  \map D\circ \map C  \, .
\end{align}
\end{prop} 
{\bf Proof.}   Suppose that $\map C$ induces a faithful causal-effect relation.  Then, let $\map R  \in \CPTP(B\to A)$ be the process that inverts $\map C$, namely $\map R\circ \map C  =  \map I_A$.  Then, for every system $B'$ and every process  $\map E  \in \CPTP(A\to B')$, one has $\map E   =       \map D  \circ \map C$, with $   \map D:  =    \map E  \circ \map R$.    

Conversely, suppose that, for every quantum system $B'$ and for every process $\map E \in \CPTP(A\to B')$,  there exists a process $\map D \in \CPTP(B\to B') $ such that  $\map E  =  \map D \circ \map C$.  In particular, one can pick $B'  \equiv A$ and $\map E  \equiv  \map I_A$, in which case the condition  $\map E  =  \map D \circ \map C$  implies that $\map C$ is correctable, and  therefore  induces a faithful cause-effect relation.  \qed

\medskip

 \medskip

In general, the presence of a faithful cause-effect relation from  $A$ to $B$ does not imply that the causal influences of $A$ propagate {\em exclusively} through $B$.  For example, quantum secret sharing  protocols, such as those presented in Ref. \cite{cleve1999share},  provide examples of processes where a given cause can have multiple causal intermediaries.     The situation is much simpler when the cause-effect relation is {\em reversible}, in the following sense:

 \begin{defi}  We say that a process $\map C\in \CPTP(A\to B)$  {\em induces  a  reversible   cause-effect relation} between $A$ and $B$ if and only if   $\map C$ is reversible,    meaning that there exists another process $\map R\in \CPTP(B\to A) $ such that 
\begin{align}  \map R \circ \map C  =  \map I_A  \qquad {\rm and} \qquad \map C\circ \map R    =  \map I_A\,,
\end{align} where $\map I_X  \in  \CPTP  (X\to X)$  is the identity process on system $X  \in  \{A,B\}$. 
\end{defi}  

In this paper, we will focus on situations where the cause-effect relation between two variables is reversible. In this case, the presence of a  cause-effect relation between variables $A$ and $B$ rules out the possibility of any cause-effect relation from $A$  to any other variable $B'$  that is independent of $B$. More precisely,  if a process  $\map D  \in  \CPTP  (A\to B\otimes B')$ is such that its reduced process $\map C  = (\map I_B\otimes  \tr_{B'}) \circ  \map D$ is reversible, then  $\map D  $ must be of the form $\map D  =  \map C\otimes \beta$, where $\beta \in  \St (B')$ is some fixed state of system $B'$.

     \subsection{Discrimination of  causal hypotheses}\label{subsect:example}   Consider the situation where an experimenter has access to a black box, implementing a quantum process with a given set of input systems and a given set of output systems.    
The goal of the experimenter is to figure out the causal relations among the systems involved in the process.  For example,  the black box could implement a process  with input system  $A$ and output system $B$, and  
  the experimenter may want to figure out whether the process induces a cause-effect relation. 
   
  In general, we will assume  that the black box is guaranteed to satisfy one, and only one, of $k$ possible  hypotheses  $({\sf H}_i)_{i=1}^k$ about the cause-effect relations occurring between its inputs and outputs. 
No further information about the process is available to the experimenter.  To figure out which hypothesis is correct, the experimenter will set up a test designed to probe the causal relations.

We will use the term {\em causal hypotheses}  for hypotheses on the causal relations induced by   a given process.  The problem of distinguishing between alternative causal hypotheses will be called {\em discrimination of causal hypotheses}.  To illustrate the problem, we will  focus on one basic instance: discover which of two variables $B$ and $C$ is the effect of a given variable $A$.  More specifically, we consider the following alternative hypotheses: 
\begin{enumerate}
\item[${\sf H}_1$:] $B$ is a causal intermediary for $A$, and $C$ is uniformly random,
\item[${\sf H}_2$:]  $C$ is a causal intermediary for $A$, and $B$ is uniformly random.
\end{enumerate}   
For simplicity, we will assume that systems $A, B$, and $C$ have the same dimension, equal to $d<\infty$.  
In terms of the process $\map C\in \CPTP(A\to (B\otimes C))$ the two hypotheses correspond to the following statements: 
\begin{enumerate}
\item[${\sf H}_1$:] $\map C$ has the form $\map C (\cdot)  =  U\cdot U^\dag\otimes  I/d$, for some unitary operator $U \in  B (\C^d)$,
\item[${\sf H}_2$:]  $\map C$ has the form $\map C (\cdot)   =   I/d  \otimes V\cdot V^\dag$, for some unitary operator $V \in  B (\C^d)$.
\end{enumerate}   
In either case, the unitary operators $U$ and $V$ are unknown to the experimenter. 

In general, every  hypothesis  on the causal relations between inputs and outputs is in one-to-one correspondence with   a  subset of CPTP maps.  
  In the above example, the two hypotheses ${\sf H}_1$ and ${\sf H}_2$ correspond to the sets 
  \begin{align}
 {\sf H}_1   & =   \Big \{  U\cdot U^\dag\otimes  I/d ~|~  U \in  B (\C^d) \, ,   \,  U^\dag U  =  U U^\dag  =  I   \Big\}   \label{hyp1} \\
 {\sf H}_2   & =   \left \{   I/d  \otimes V\cdot V^\dag~|~  V\in  B(\C^d)  \,  ,  V^\dag V=  V V^\dag   =  I \right\} \, . \label{hyp2}
  \end{align}
  The problem is to determine whether the given process $\map C$ belongs to $\sf H_1$ or to $\sf H_2$.    
  
\begin{figure}[htbp!]

\centering
\includegraphics[width=100mm]{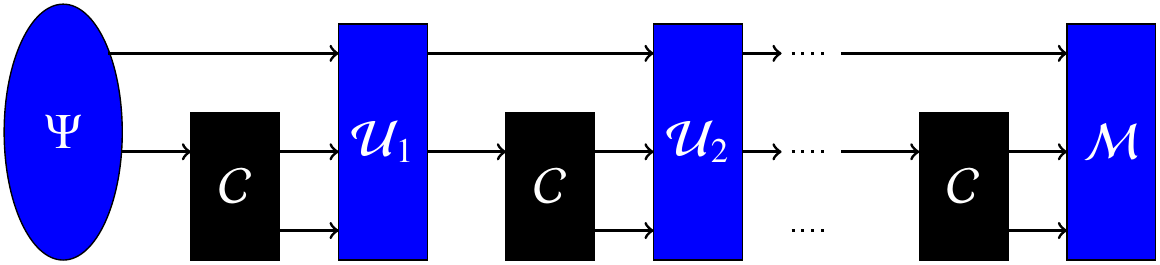}
\caption{General setup for testing causal hypotheses on a process $\map C$ with one input variable and two output variables \cite{Chiribella-circuit2008, Chiribella-memory2008, chiribella-dariano-2009-pra}.  The experimenter prepares an initial state $\Psi$, possibly entangled with an auxiliary system,  and probes the process for $N$ times. Between one execution of the process $\map C$ and the next, the experimenter can intervene by performing a unitary gate $\map {U}_i$ on the output systems of $\map C$ and on some auxiliary  system in the laboratory.    Finally, a measurement $\map M$ is performed and the outcome is used to produce a guess of the correct causal hypothesis.}
\label{fig:schematic}
\end{figure}

For the discrimination between these two hypotheses we will consider  setups that probe  the unknown process $\map C$  for $N$ times, by inserting them in a sequential quantum circuit, as illustrated in Figure \ref{fig:schematic}.  
 As a figure of merit, we will consider the minimization of the probability of error, in the worst case over all possible processes $\map C$ that are compatible with the causal hypotheses (\ref{hyp1}) and (\ref{hyp2}).

In general, the error probability in distinguishing two known quantum processes decays exponentially with the number of interrogations \cite{yu2017chernoff}. Informally, this means that the error probability scales asymptotically as $p_{\rm err}  (N)  \approx  2^{  -  N \, R}$ for some exponent $R  >0$, called the {\em decay rate}.  More formally,   the decay rate is defined as  $R  :  =  \liminf_{N\to \infty}  -\log [p_{\er}(N)]  /N $, where we assume base $2$ for the logarithm here and in the rest of the paper.  
In the following, we will use the decay rate to assess the performance of various strategies for  distinguishing causal hypotheses.

\section{Optimal classical strategy for finding a causal intermediary}
\label{sec:opt-class}
Let us consider  the classical version of the problem in Subsection \ref{subsect:example}.  The classical version involves  three classical random variables $A,B,$ and $C$, each with sample space of cardinality  $d<\infty$.  
The two causal hypotheses are:  
\begin{enumerate}
\item[${\sf H}_1$:]    $B$ is a permutation of $A$,  and $C$ is uniformly random, 
\item[${\sf H}_2$:]  $C$ is a permutation of $A$, and $B$ is uniformly random. 
\end{enumerate}
In either case,  the exact form of the permutation is unknown to the experimenter.

For parallel strategies where the unknown process is probed on $N$ independent inputs, 
 the optimal error probability $p_\er^{\rm C} (N)$  is \cite{Giulio2019}
 \begin{align}\label{perC}
p_\er^{\rm C} (N)=\frac{1}{2 d^{N-1}}.
\end{align} 
This implies the optimal decay rate  is 
\begin{align}
R_{\rm C}=\log d.
\end{align}
A result by Hayashi \cite{hayashi2009discrimination} implies that the decay rate cannot be improved by considering more general sequential strategies such as those in Figure \ref{fig:schematic}.

\section{Quantum advantages}
\label{sec:opt-quant}

Classical random variables can be regarded as quantum systems that have lost coherence with respect to a privileged orthonormal basis of ``classical states".   But what if the coherence is preserved?  In the following we will show that the possibility of probing processes through coherent superpositions of classical inputs can enhance our ability to identify the correct causal hypothesis.

\subsection{A benefit of quantum coherence}

The three classical random variables $A, B$, and $C$ considered in the previous section can be considered as the decohered versions of three quantum systems $A$, $B$, and $C$ with  $d$-dimensional Hilbert spaces.
Similarly, a permutation of the values of the random variable $A$ according to an element $\pi $ of the symmetric group $\grp S (d) $ can be regarded as the decohered version of the unitary permutation operator $U_\pi =  \sum_{i=1}^d   |\pi(i)  \>\<i|$, where $\{ |i\>  \}_{i=1}^d$ is the orthonormal basis representing the classical states. 

 In this scenario, the two ``classical'' causal hypotheses become 
 \begin{itemize}
 \item[${\sf H}_1$:] $\map C$ has the form $\map C (\cdot)  =  U_\pi\cdot U_\pi^\dag\otimes  I/d$, for some permutation $\pi\in  \grp{S}  (d)$,
 \item[${\sf H}_2$:]  $\map C$ has the form $\map C (\cdot)   =   I/d  \otimes U_\sigma\cdot U_\sigma^\dag$, for some permutation $\sigma \in  \grp{S} (d)$.
\end{itemize}
  
To distinguish between these two hypotheses, the experimenter could prepare $N$ probes in the superposition state
\begin{equation}
\ket{e_0}=\frac{1}{\sqrt d}\sum_{i=0}^{d-1} \ket{i} \, ,
\end{equation}
which is invariant under permutations. 
   Thus, the unknown process $\map{C}$ yields either $\left(\ket{e_0}\bra{e_0}\otimes\frac{\mathbb{I}}{d}\right)^{\otimes N}$ or  $\left(\frac{\mathbb{I}}{d}\otimes\ket{e_0}\bra{e_0}\right)^{\otimes N}$ depending  on which  causal hypothesis is correct.

The probability of error can be computed by taking advantage of  the symmetry of the problem, which implies that the worst case error probability is equal to the average error probability when the prior probability for the two hypotheses $\sf{H}_1$ and ${\sf H_2}$ is uniform.  In this case,  Helstrom's theorem \cite{helstrom1969quantum} states that the minimum  error probability in distinguishing between the states $\rho_1$ and $\rho_2$, given with uniform {\em a priori} probability, is
\begin{align}
p_{\er, {\rm ave}}   (\rho_1, \rho_2)   =    \frac {  1  -  \frac 12  \left\|  \rho_1  - \rho_2\right\|_1}2  \, ,
\end{align}
where $\|  G \|_1  :  =  \tr  \sqrt {G^\dag G}$ denotes the trace norm of a generic trace class operator $G \in T(\spc H)$. 
Applied to our problem, Helstrom's theorem yields the error probability 
\begin{align}
\label{eq:coh-ad}
p_{\er}^{\coh}= \frac{1}{2d^N}.
\end{align}
By comparison with the classical error probability (\ref{perC}), we can see that coherence provides a reduction of the error probability by a factor $d$.    This is a relatively minor improvement,  as it does not affect the decay rate. Still, it is an interesting first illustration of how quantum effects can affect the discrimination of causal hypotheses.

\subsection{General quantum scenario}\label{subsec:noreference}
Let us move on now to the general quantum scenario, where the relation between cause and effect is described by an arbitrary unitary operator (not necessarily a permutation operator, as in the previous subsection). 

At first, this problem appears to be much more challenging for our experimenter, since the dependence between cause and effect can be any arbitrary  element of the special unitary group $\grp {SU} (d)$.   Surprisingly, however, the {\em same} error probability as in  Eq. \eqref{eq:coh-ad} can be achieved.  

Let us consider the case where $N$ is an integer multiple of $d$, say $N=  d \, t$ for some non-negative integer $t$.  In this case, a universal quantum strategy with error probability \eqref{eq:coh-ad}  is to divide the $N$ probes into groups of $d$, and to prepare each group of probes in $\grp {SU} (d)$ singlet state
\begin{align}
\ket{S_d}=\frac{1}{\sqrt{d!}}\sum_{k_1, \cdots, k_d} \epsilon_{k_1\cdots k_d} \ket{k_1,\cdots,k_d},
\end{align}
where $\epsilon_{k_1\cdots k_d}$ represents the totally antisymmetric tensor and the summation extends over all the vectors in the computational basis. 

Each of the $d$ particles in each group is then fed into one use of the unknown process. Since the singlet state is invariant under unitary transformations, the problem becomes to distinguish between the state $|S_d\>\<S_d|^{\otimes t} \otimes (I/d)^{\otimes N} $  (hypothesis $\sf{H}_1$) and $(I/d)^{\otimes N}  \otimes |S_d\>\<S_d|^{\otimes t}  $ (hypothesis $\sf{H}_2)$.    In this  case, Helstrom's theorem again gives error probability    
 $1/(2d^N)$, which is equal to the value $p_{\er}^{\coh}$ obtained in the previous subsection.   
 
 If $N$ is not a multiple of $d$, a rough strategy is to use only the first $\widetilde N  : =  d\,  \lfloor N/d \rfloor$  probes, and to  apply the above procedure. In this way, one obtains error probability $1/(2d^{\widetilde N})$, which is suboptimal but still has the same decay rate as the coherent strategy we saw in the previous subsection.

\subsection{Parallel strategies with  an external reference system}\label{subsec:reference}
We now show that the decay rate of the minimal error probability can be  doubled if the experimenter uses  an additional reference system entangled with $N$ input probes in the protocol. For simplicity, we will assume that $N$ is a multiple of $d$.

The basic idea for constructing the quantum-enhanced strategy is the following.  In the absence of a reference system, the strategy was to divide the $N$ probes into $N/d$ subgroups of size $d$, and to put the particles in each group in the singlet state.    However, there are many ways of dividing $N$ particles into groups of $d$.  Each of these ways leads to the error probability $p_{\er}^{\coh}  =  1/2d^N$.  
What about trying all possible ways in a coherent superposition?  

Consider an external reference system with an orthonormal basis $\{\ket{i}\}_{i=1}^{G_{N,d}}$, where the index $i$ labels the possible ways in which $N$ particles can be divided into groups of $d$, and  $G_{N,d}$ denotes the number of such ways.  
 Then, one can construct the superposition state
\begin{align}\label{quantumsup}
\ket{\psi}= \frac{1}{\sqrt{G_{N,d}}}\sum_{i=1}^{G_{N,d}} \left(\ket{S_d}^{\otimes {N/d}}\right)_i\otimes\ket{i} \,,
\end{align} 
where $\left(\ket{S_d}^{\otimes {N/d}}\right)_i$ denotes the product of $N/d$ singlet states distributed in the $i$-th way.

While a classical randomization  over all the possible ways to group the $N$ particles is useless, the  quantum superposition (\ref{quantumsup})  turns out to be {\em very} useful. After some algebra \cite{Giulio2019}, it is possible to show that the optimal setup using the state (\ref{quantumsup}) has  error probability 
\begin{align}
p_{\rm err}^{\rm Q}(N)=\frac{m(N,d)}{2d^N}\left( 1-\sqrt{1-\frac{1}{m(N,d)^2}}  \right),
\end{align}
where 
\begin{align}\label{mnd}
m  (  N,d)   :=    N!  \,       \prod_{i=1}^d      \frac{(d-i)!}{(N/d+d-1)!}  
\end{align} is the multiplicity of the trivial representation of $\grp{SU} (d)$ in the $N$-fold tensor representation  $U\mapsto U^{\otimes N}$.
For large $N$, the above expression can be approximated as 
\begin{align}
p_{\rm err}^{\rm Q}(N)\approx \frac{1}{4 m(N,d) \, d^N}\, .
\end{align}
 Using the explicit expression (\ref{mnd}), we then obtain the decay rate
\begin{align}
R_{\rm Q} = -\lim_{N\rightarrow\infty} \frac{\log  [ p_{\er}^{\rm Q}(N)]}{N} = 2 \log (d)= 2 R_{\rm C} \,.
\end{align}
Thus, the presence of entanglement between the probes and the external reference system allows one to  double the decay rate of the error probability. 
 
 The asymptotic limit can be achieved for sufficiently small number of interrogations of the unknown process. For example,  an error probability less than $10^{-6}$ can be achieved by using $12$ interrogations for discrimination of a causal relation between two quantum bits, while $20$ interrogations are necessary for classical binary variables in order to achieve the same error threshold.

\subsection{The ultimate quantum limit}
In the previous subsections (\ref{subsec:noreference}) and (\ref{subsec:reference}), we  considered strategies where the unknown process was applied for $N$ times in parallel on $N$  input systems.  These strategies are a special case of the sequential strategies shown in Figure \ref{fig:schematic}.  
Can our experimenter further reduce the error probability by using these more general strategies? 

A useful tool to address this question is the notion of {\em fidelity divergence} between two processes, introduced in Ref. \cite{Giulio2019}, and later generalized to a broader set of channel divergences in Ref. \cite{berta2018amortized}.  The fidelity divergence between processes  $\map C_1$ and $\map C_2$ is defined as 
\begin{align}
\label{fid}
\partial F  (\map C_1, \map C_2)  := \inf_R \inf_{\rho_1, \rho_2} \frac  {  F \left( \rho_1',\rho_2' \right)}{F( \rho_1,\rho_2)},
\end{align}  
where   $\rho_1$ and $\rho_2$ are states of the process input and of a reference system $R$,  $\rho_1':=(\map C_1\otimes \map I_R)(\rho_1)$, and $\rho_2':= (\map C_2\otimes \map I_R)  (\rho_2)$. In order for the above expression to be well defined, the infimum is taken over all the states $( \rho_1,\rho_2)$ such that $F( \rho_1,\rho_2)\neq 0$. 

Let us denote by $p^{\seq}_{\er}  (\map C_1,\map C_2; N)$  the error probability  in distinguishing between the two processes $\map C_1$ and $\map C_2$ using a sequence of $N$ interrogations, as in Figure \ref{fig:schematic}. 
 It can be shown that the error probability satisfies the bound \cite{Giulio2019}
\begin{align}
p^{\seq}_{\er}  (\map C_1,\map C_2; N)  \ge  \frac {\partial F   (\map C_1, \map C_2)^N  }4.
\end{align}
In the special case where  $\map{C}_1 = \map{U} \otimes \mathbb{I}/d$ and $\map C_2  = \mathbb{I}/d \otimes \map{V}$ with $\map{U} (\cdot)  =  U \cdot U^\dag$ and $\map V(\cdot)  =  V\cdot V^\dag$  being  fixed unitary processs, the fidelity divergence is $1/d^{2}$, and  the error probability is lower bounded as 
\begin{align}
p^{\seq}_{\er}  (\map C_1,\map C_2; N)  \ge  \frac {1}{4d^{2N}}.
\end{align}
Hence, the decay rate cannot be larger than    $2\log d$, {\em even if the unitaries $U$ and $V$ are known!}  
If the unitaries $U$ and $V$ are unknown, as in the causal discrimination scenario, then the decay rate can only be   $2\log d$ or smaller.  
This observation proves that the decay rate $2\log d$, achievable with a parallel strategy, is optimal among all decay rates achievable by arbitrary sequential strategies.

As a further curiosity, one may ask  whether the rate could be improved in some
 exotic scenario  where the order of the $N$ interrogations is  indefinite, unlike in the scenario of Figure \ref{fig:schematic}, where the $N$  interrogations  happen in a well-defined sequential order. 
 In principle, quantum probability theory is logically compatible with scenarios where  quantum processs are combined in an indefinite causal order
\cite{chiribella2009beyond,oreshkov2012quantum,chiribella2013quantum}. Physically, these scenarios may arise in exotic quantum gravity regimes, although research on such realizations is still in its infancy (and, of course,  no complete theory of quantum gravity has been formulated yet).  
 Still, as a theoretical possibility, one can already investigate the question of whether  
  the ability to test  a process for $N$ times in an indefinite causal order could help identifying the causal relations occurring between its inputs and outputs. 
   
For the identification of the causal intermediary, the answer turns out to be negative. The proof strategy is to bound the discrimination error for two simple processes, namely, $\map C_1  =  \map I\otimes \mathbb{I}/d$ and $\map C_2  = \mathbb{I}/d \otimes \map I$ using arbitrary setups with indefinite order. 
   Using semidefinite programming, Ref. \cite{Giulio2019} showed that  the minimum error probability over all setups that place $N$ uses of the unknown process in an indefinite order is 
\begin{align}
p_{\er}^{\ind} (\map C_1,\map C_2; N)  \geq  \frac{1}{2} \left(1- \sqrt{ 1-  \frac 1 {d^{2N}}}\right). 
\end{align}
 This result establishes  the decay rate $2\log d$ as the ultimate limit placed by quantum mechanics to the identification of a causal intermediary.  In addition,  the strong duality of semidefinite programming guarantees that there exists a suitable setup (possibly  requiring indefinite causal order) that achieves the above  error probability exactly.

\section{Other examples of speedups in the identification of causal hypotheses}
\label{sec:apps}
\subsection{Multiple candidates for the causal intermediary}
In the previous sections, we have seen how to identify the causal intermediary of a variable $A$ among two possible candidates $B$ and $C$. What about more than two candidates? 
For $k>2$ candidates, the derivation is technically more involved, but the main results remain unchanged: 
 
\begin{itemize}
\item In the classical case, when causal relations are described by arbitrary permutations, the minimal error probability is given by
\begin{align}
p_{\er, k}^{\rm C}=\frac{(k-1)}{2d^{N-1}} +O(d^{-2N}) \, ,
\end{align}
and the decay rate is $\log d$.  

\item  In the quantum case, parallel strategies without a reference system achieve error probability    
\begin{align}
p_{\er, k}^{\rm Q}=\frac{(k-1)}{2d^{N}} +O(d^{-2N}) \, ,
\end{align}
and the decay rate is still $\log d$. In contrast, parallel strategies using entanglement with an external reference system can achieve a doubled decay rate $2\log d$. 
\end{itemize}

\subsection{Detection of causal link between two variables}  A basic example of identification of causal hypotheses is to determine whether there is a causal link between two variables $A$ and $B$.  In this case, the task is  to determine whether $B$ is a causal intermediary for $A$, or whether $B$  fluctuates at random  independently of $A$. As it turns out, entanglement with a reference system can  once again double the decay rate of the error probability:   the classical error probability decays with rate $\log d$,  while the quantum error probability with reference systems decays with doubled rate $2\log d$.

\subsection{Identification of the  cause of a variable} 

Another interesting problem is  to identify which variable in a given set  $\{ A_1,\dotsm  A_m\}$
is the cause for a given variable $B$ (assuming that one and only one variable  can be the cause).  Again, we assume that causal relation is induced by a reversible process (a permutation in the classical case, or a unitary gate in the quantum case) and that all the variables have sample space of cardinality $d$ in the classical case, or Hilbert space of dimension $d$ in the quantum case.   

Classically, the  problem is to find the random  variable $A_i$  such that $B$ is a permutation  of $A_i$, with $i \in  \{1,\cdots, m\}$. In the simplest case, when the permutation is known,  the cause can be determined without error by interrogating the unknown process  $\ceil{\log_d(m)}$ times.

In the quantum case, Ref. \cite{Giulio2019} showed that, if the unitary operator inducing the causal relation is known,  then there exists a tests that achieves  error probability
\begin{align}
p_{\er}(N)   =  \frac{m-1}{d^{2N} + m-1}
\end{align}
This means that one can get an error probability smaller than any desired  $\epsilon>0$ by using  $N=\ceil{(1+\epsilon) (\log_d m) /2}$  interrogations, which is approximately half of the number of classical interrogations. 

In the case where the dependency between cause and effect is unknown (arbitrary permutation in the classical case, or arbitrary unitary operator in the quantum case),  the analysis is more complex. However, it is still possible to show that quantum strategies can identify the cause using only  $N=\ceil{(1+\epsilon) (\log_d m) /2}$ interrogations.

\section{Discussion and conclusions}
\label{sec:conc}
In this paper, we reviewed the framework and the results of Ref. \cite{Giulio2019}, which showed that quantum features such as coherence and entanglement offer advantages in detecting cause-effect relations induced by reversible processes. 

In the problem of identifying the effect of a given variable,  it was shown that entanglement between the probes and an external reference system can double the rate at which the error probability decays.  For classical random variables with 
$d$ possible values, the decay rate is  $\log d$.   
For quantum systems of dimension $d$, the decay rate is $2 \log d$.   As it turns out, the value $2 \log d$  is the ultimate limit posed by quantum  theory to the problem of identifying the causal intermediary of a given variable.

Interestingly,  both the classical and quantum decay rates can be expressed as $\log \dim \St_\R (A)$ where $\St_\R  (A)$ is the vector space spanned by linear combinations of states of system $A$.  For classical systems, the states are probability distributions over the sample space, and the dimension of the corresponding vector space is $d$. For quantum systems, the states are density matrices, and the dimension of the vector space is $d^2$.  It would be interesting to study the problem of causal hypothesis discrimination in  toy theories with higher dimensional state spaces,  such as quantum theory on quaternionic Hilbert spaces \cite{barnum2015some} or the quartic toy  theory proposed by \. Zyczkowski in  Ref. \cite{zyczkowski2008quartic}.

On a more practical side, it is important to extend the analysis  from the idealized scenario where the cause-effect dependencies are induced by reversible processes, to the more realistic scenario where they are induced by general noisy processes.   Preliminary results in Ref. \cite{Giulio2019} indicate that quantum advantages may still persist for sufficiently low noise levels. However, a fully general treatment of noise  is still lacking and  will be important  for future applications.

Given the success of causal discovery algorithms in classical statistics and machine learning,  it is natural to expect that the development of quantum causal discovery algorithms may have  applications to the burgeoning field of quantum machine learning \cite{schuld2015introduction,biamonte2017quantum,dunjko2018machine}. This connection is largely unexplored and represents an exciting direction of future research.

\medskip

{\bf Acknowledgements.}   GC is grateful to R Spekkens,  D Schmid, and M T Quintino  for stimulating discussions on the notion of causality for quantum processes.  This work is supported by the National Science
Foundation of China through Grant No. 11675136, by Hong Kong Research Grant Council
through Grants No. 17326616 and 17300918, by the Croucher Foundation, and by the John Templeton Foundation through grant  61466, The Quantum Information Structure of Spacetime  (qiss.fr).
The opinions expressed in this publication are those of the authors and do not necessarily reflect the views of the John Templeton Foundation. Research at the Perimeter Institute is supported by the Government of Canada through the Department of Innovation, Science and Economic Development Canada and by the Province of Ontario through the Ministry of Research, Innovation and Science.

    \bibliographystyle{unsrt}

  \bibliography{cause-effect}

\end{document}